\documentstyle[12pt]{article}
\newcommand{\be}{\begin{equation}}
\newcommand{\bea}{\begin{eqnarray}}
\newcommand{\eea}{\end{eqnarray}}
\newcommand{\ba}{\begin{array}}
\newcommand{\ea}{\end{array}}
\newcommand{\ee}{\end{equation}}

\expandafter\ifx\csname mathbbm\endcsname\relax

\else

\fi
\def\o{\over}

\begin{document}
\begin{titlepage}
\hfill
\vbox{
    \halign{#\hfil         \cr
           %hep-th/9710020 \cr
           IPM-97-238   \cr
           2 Oct. 1997   \cr
           } % end of \halign
      }  % end of \vbox
\vspace*{3mm}
\begin{center}
{\LARGE $N=(4,4)$ 2D Supersymmetric Gauge Theory and Brane Configuration \\}
\vspace*{20mm}
{\ Mohsen Alishahiha \footnote{e-mail:alishah@theory.ipm.ac.ir}}\\
\vspace*{1mm}
{\it Institute for Studies in Theoretical Physics and Mathematics, \\
 P.O.Box 19395-1795, Tehran, Iran } \\
%{\it Department of Physics, Sharif University of Technology, \\
%\it  P.O.Box 11365-9161, Tehran, Iran }\\
\vspace*{25mm}
%\maketitle
\end{center}
\begin{abstract}
We construct type II A brane configuration of $N=(4,4)$ supersymmetric 
two dimensional gauge theory with gauge group $U(1)$ and $N_f$ 
hypermultiplets in the fundamental representation. 
By lifting to M-theory (strong coupling), we can see the origin of the 
R-symmetry enhancement of the Coulomb branch. One can also
find two theories which become equivalent at strong coupling.
\end{abstract}
\end{titlepage}
\newpage
Brane theory gives us a very useful tool to construct supersymmetric gauge 
theory in various dimension with various supercharges. At first, Hanany
and Witten \cite{HW} constructed a particular brane configuration in 
type II B
string theory which describes mirror symmetry in three dimensional 
supersymmetric gauge theory with 8 supercharges. It was shown that there
is a duality between the Higgs branch and the Coulomb branch of $N=4$ SYM in
three dimension\cite{IS}. This duality acts as mirror symmetry, exchanging 
the Higgs and Coulomb branches of the theories. Applying string duality
to the configuration of branes in type II B string theory, one can provide
an explanation for the mirror symmetry.

It has been shown \cite{EGK} by a particular configuration of type II A 
string
theory that one can obtain supersymmetric $U(n)$ gauge theory in four 
dimension
with four supercharges. Then by making certain deformation in brane 
configuration, Seiberg's duality can be realized in brane  theory.
Introducing orientifold plane in brane configuration help us to generalize
this work to other classical Lie gauge groups. \cite{SO}

$N=2$ four dimensional gauge theory with gauge group $SU(n)$ was also 
obtained from brane configuration in type II A string theory\cite{WIT}.
By lifting from type II A to M-theory, one can obtain the exact solution
of $N=2$ $D=4$ SYM theory. More precisely, in M-theory point of view, we
have a five brane with worldvolume $R^{3,1} \times \Sigma$, where the 
theory on the $R^{3,1}$ space is $N=2$ $D=4$ SYM theory and $\Sigma$ is the
Seiberg-Witten curve corresponding to it. This work was generalized to other
classical Lie gauge groups in \cite{GSU}.

There is another method to studying gauge theory from string theory, 
the so called Geometric Engineering. By this approach the same class of the 
above theories can be studied by using wrapped D-branes on Calabi-Yau cycles
in type II B and A string theory compactified on the K3-fibered Calabi-Yau
3-fold.\cite{LV} This approach was generalized in \cite{KVA} to explain mirror
symmetry in $N=4$ three dimensional gauge theory as well as Seiberg-Witten
models with product of gauge groups.

Recently, Hanany and Hori \cite{HH}
used brane configuration in type II A string theory
to construct $N=2$ supersymmetric gauge theory in two dimension. The aim of 
this article is to construct brane configuration for $N=(4,4)$ gauge theory
in two dimension, through a particular brane in type II A and M-theory.
Very recently, this theory was analyzed in \cite{WIT2} and \cite{DS}. 

Here we will use three kinds of objects in type II A string theory: 
NS 5-brane with worldvolume $(x^0,x^1,x^2,x^3,x^4,x^5)$ which lives
at a point in the $(x^6,x^7,x^8,x^9)$ directions. D 2-brane with worldvolume
$(x^0,x^1,x^6)$ at the point $(x^2,x^3,x^4,x^5,x^7,x^8,x^9)$ and also
D 4-brane with worldvolume $(x^0,x^1,x^7,x^8,x^9)$ living at  the point 
$(x^2,x^3,x^4$, $x^5)$ and $x^6$. 

Brane configuration which we will consider is two NS 5-branes at 
$r_1=(x^7_1,x^8_1,x^9_1)$, $x^6=0$ and $r_2=(x^7_2,x^8_2,x^9_2), x^6=L$.
A D 2-brane suspended between these two NS 5-branes, so it is finite in 
the $x^6$ direction, and $N_f$ D 4-branes at 
$m_i=(x^2_i,x^3_i,x^4_i,x^5_i)$ between two NS 5-branes.( See 
the following figure) 
\vspace*{1mm} 
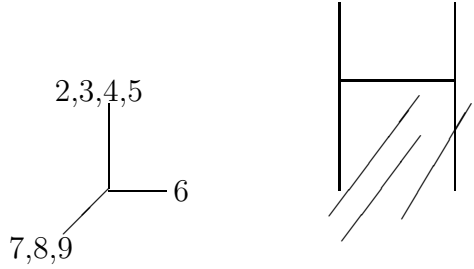
\begin{center}
%TexCad Options
%\grade{\off}
%\emlines{\off}
%\beziermacro{\on}
%\reduce{\on}
%\snapping{\off}
%\quality{2.00}
%\graddiff{0.01}
%\snapasp{1}
%\zoom{1.00}
\unitlength 1.00mm
\linethickness{0.4pt}
\begin{picture}(60.20,45.00)
\put(12.00,31.67){\line(0,-1){11.33}}
\put(12.00,20.33){\line(-1,-1){6.00}}
\put(12.00,20.00){\line(1,0){7.67}}
\put(10.67,33.00){\makebox(0,0)[cc]{2,3,4,5}}
\put(3.00,12.00){\makebox(0,0)[cc]{7,8,9}}
\put(21.67,20.00){\makebox(0,0)[cc]{6}}
\put(42.67,20.00){\line(0,1){25.00}}
\put(58.00,45.00){\line(0,-1){25.00}}
\put(41.33,16.67){\line(3,4){12.00}}
\put(43.00,13.33){\line(3,4){10.50}}
\put(51.00,16.33){\line(3,5){9.20}}
\put(43.00,34.67){\line(1,0){15.00}}
\end{picture}
\end{center}
\vspace*{.3mm} 

It is easy to see that these configurations preserve 8 
supercharges in intersection worldvolume $(x^0,x^1)$, so we have 
$N=(4,4)$ gauge theory in two dimension. This brane configuration can be 
obtained by T-duality from the configuration of \cite{HW} in direction $x^2$.
So it is natural to expect that some properties of that configuration may
induce to our theory. For example in brane configuration of three dimensional
$N=4$ gauge theory the $x^6$ coordinate of D 5-brane (matter) appears to be
irrelevant, so in our case we expect that $x^6$ is irrelevant too. This
fact as we will see leads us to an interesting equivalence between theories. 

The presence of all these branes break the Lorentz group $SO(1,9)$ to
1+1 Lorentz group with $SO(4)\times SU(2)_R$ global symmetry. $SO(4)$ 
corresponds to rotation in $x^2,x^3,x^4,x^5$ directions which we write as
$SU(2)\times SU(2)$. $SU(2)_R$ is rotation in $x^7,x^8,x^9$ directions.

The two NS 5-brane are at $r_1$ and $r_2$, $r=r_2-r_1$ is
Fayet-Iliopoules terms, so it is in $(1,1,3)$ representation of $SO(4)\times
SU(2)_R$, as expected\cite{DS}, where the representation of $SO(4)$ are 
labeled by $SU(2)\times SU(2)$. When $r=0$, the D 2-brane can suspend 
between two NS 5-branes and 
preserve 8 supercharges in 1+1 dimension. In this case we have $U(1)$ gauge
theory which is in the Coulomb phase and can be parametrized by scalars 
in vector
multiplet. In brane language, these scalars are fluctuations of D 2-brane in
$x^2,x^3,x^4,x^5$ directions which we set $u=x^2+ix^3, v=x^4+ix^5$. They 
transform as $(2,2,1)$ of $SO(4)\times SU(2)_R$.

Now consider $N_f$ D 4-branes between the two NS 5-branes. The position of these
D 4-branes, $m_i$, are bare masses of hypermultiplets which transform as  
$(2,2,1)$ under global symmetry $SO(4)\times SU(2)_R$. Strings stretched
between D 2-brane and D 4-branes give us hypermultiplets in the fundamental 
representation. If two of $m_i$'s become equal, D 2-brane can break and 
suspends between these two D 4-branes, and the theory is in the Higgs phase. 
Along the Higgs branch there are scalars transforming in $3+1$ of $SU(2)_R$.
In brane language, these scalars correspond to fluctuation of D 2-brane in 
$x^7,x^8,x^9$ directions. If we look at this configuration from the point 
of view of M-theory the last one which is singlet under $SU(2)_R$, 
becomes manifest.
The position of NS 5-branes in $x^{10}$ (compact direction in M-theory) 
can be interpreted as $\theta$ angle ($\theta=x^{10}_2-x^{10}_1$)

In the case of $r=0$ (and $\theta=0$ in quantum level \cite{SOWI} which
means M-theory level) the theory can have the Coulomb branch. If $r\neq0$
(or $\theta \neq 0$) the theory can not have the Coulomb branch, which means
that if we want to have a supersymmetric configuration, the D 2-beane 
must be broken into D 2-branes between NS 5-branes and D 4-branes. 
If the D 4-branes have the same position in $x^2,x^3,x^4,x^5$ 
(equal mass $m_i=m_j$), the  theory is in the
Higgs branch. Note that, for $N_f=1$ the D 2-brane must be suspended between
D 4-brane and two NS 5-branes, so thses two D 2-brane are fixed from two 
sides by boundary condition, then the theory does not have the Coulomb 
and Higgs branches, classically \cite{DS}. (following figure) 
\vspace*{1mm} 
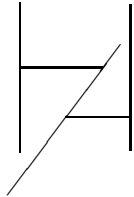
\begin{center} 
%TexCad Options
%\grade{\off}
%\emlines{\off}
%\beziermacro{\on}
%\reduce{\on}
%\snapping{\off}
%\quality{2.00}
%\graddiff{0.01}
%\snapasp{1}
%\zoom{1.00}
\unitlength 1mm
\linethickness{0.4pt}
\begin{picture}(35.00,40.00)
\put(20.33,20.00){\line(0,1){20.00}}
\put(35.00,20.33){\line(0,1){19.33}}
\put(18.67,14.33){\line(3,4){15.00}}
\put(35.00,24.67){\line(-1,0){8.67}}
\put(20.33,31.33){\line(1,0){11.00}}
\end{picture}
\end{center} 
\vspace*{.3mm} 

It was shown \cite{WIT2} that quantum mechanically 
the theory with $N_f=1$  can have Higgs branch although classically it 
does not have it.

The distance between two NS 5-branes determines the gauge coupling constant 
of the two dimensional theory, more precisely
\be
{1 \o g_2^2}={L \o \lambda}
\ee
where $\lambda$ is the string coupling constant. The gauge coupling constant
can be promoted to a background vector superfield, thus it can affect 
only the metric on the Coulomb branch\cite{DS}. From our interpretation
of mass and Fayet-Iliopoulos termes it is easy to see that \cite{DS}: i) The
mass terms can be regarded as scalar components of vector superfield, thus
can affect only the Coulomb branch. ii) The Fayet-Iliopoulos terms can
be promoted to hypermultiplets, thus can affect the metric on the Higgs branch.

In the spirit of \cite{HW} presence of $N_f$ D 4-branes induce magnetic 
charges in $(u,v)$ space. So it can affect the metric on the Coulomb branch.
\footnote{Note that in two dimension, moduli space of vacua is not well defined.
But in the spirit of the Born-Oppenheimer approximation, on can refer to it as 
a target space on non-linear sigma model\cite{DS}}
Minimizing the total five brane worldvolume, we find four dimensional Laplace
equation which has solution of the form $ A+{B \o X^2}$. So for $U(1)$ gauge
symmetry we will have the following metric on the Coulomb branch
\be
ds^2=({1\o g^2}+{N_f \o X^2})d^2X
\ee
where $X^2=u{\bar u}+v{\bar v}$. In the case of $m_i \neq0$ we find
\be
ds^2=({1\o g^2}+{1 \o |X-m_1|^2} +{1 \o |X-m_2|^2}+\cdots+
{1 \o |X-m_{N_f}|^2}) d^2X 
\ee
which is one-loop correction to the metric in the Coulomb branch\cite{DS}.
Therefore, the generalized Kahler potential is \cite{MR}
\be
K={1\o g^2_2}(u{\bar u}-v{\bar v})-N_f\int^{{v{\bar v}\o u{\bar u}}}
{d\eta \o \eta} \ln(\eta+1) + \ln u \ln{\bar u}
\ee
This leads to the torsion
\be
B=-{1\o 4}N_f \sin^2{\theta \o 2}d\phi\wedge d\chi
\ee
where $0\leq \theta <\pi, 0\leq \phi <\pi, 0\leq \chi <4\pi$ are angular 
coordinates on the unit three sphere
\bea 
u=e^{i(\chi-\phi)/2}\cos{\theta \o 2} &\,\,\,
v=e^{i(\chi+\phi)/2}\sin{\theta \o 2}
\eea

Since the intersection of the D 2-brane and NS 5-branes is string, it is
natural to interprete the torsion as the anti-symmetric $B_{\mu\nu}$
field living on the NS 5-brane of type II A string theory.

Now, let us  lift to the M-theory brane configuration. We should also 
consider $x^{10}$ compact direction. In this case, the NS 5-branes and 
D 2-brane do not change, but D 4-brane becomes wrapped M 5-brane. So we
expect that the metric in the Coulomb branch does not give any quantum
correction.\cite{DS}

If we look at this theory at strong coupling region which is M-theory
brane configuration, one finds some interesting facts.
First, the configuration in M-theory consists of M 5-brane and membrane.
In our case the Lorentz group of 11-dimensional M-theory is broken to
$SO(1,1)\times SO(4)\times SO(4)$ by our brane configuration. This two
$SO(4)$ can be interpreted as R-symmetry. The first $SO(4)$ which acts on the
$x^2,x^3,x^4,x^5$ directions, corresponds to R-symmetry of the Higgs branch.
The second $SO(4)$ which acts on the $x^7,x^8,x^9,x^{10}$ directions, is the
R-symmetry of the Coulomb branch. Note that the R-symmetry of the Coulomb
branch of the theory enhances from $SU(2)_R$ to $SO(4)$ at strong coupling, as
conjectured in \cite{WIT2}. So in the strong coupling, the R-symmetry of
the Higgs and Coulomb branches are the same. It is very similar to three 
dimensional $N=4$ gauge theory, where we have mirror symmetry, 
which exchanges Coulomb
and Higgs branches. So we expect that there is some similar symmetry in
two dimensional $N=(4,4)$ gauge theory in strong coupling. Moreover, consider
$U(1)$ gauge theory with $N_f=2$. As we mentioned, the $x^6$ coordinate of
D 4-branes is irrelevant\cite{HW}. So we can move them in this direction 
through the NS 5-brane, then two D 2-brane are created\cite{HW}. Then 
we will have two D 2-branes between D 4-branes and NS 5-branes and also a 
D 2-brane  between NS 5-branes.  Now consider the theory to be in the 
Higgs phase. In this
case we have one D 2-brane between D 4-branes and two D 2-branes between 
NS 5-branes and D 4-branes. Now if we lift to M-theory (strong coupling),
since both NS 5-branes and D 4-branes become M 5-brane, it seems
that in strong coupling these two theories are equivalent. In M-theory
both theories have the following brane configuration.
\vspace*{1mm}  
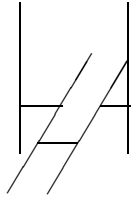
\begin{center} 
%TexCad Options
%\grade{\off}
%\emlines{\off}
%\beziermacro{\on}
%\reduce{\on}
%\snapping{\off}
%\quality{2.00}
%\graddiff{0.01}
%\snapasp{1}
%\zoom{1.00}
\unitlength 1mm
\linethickness{0.4pt}
\begin{picture}(35.00,40.00)
\put(20.33,20.00){\line(0,1){20.00}}
\put(34.67,20.00){\line(0,1){20.00}}
\put(18.67,14.67){\line(3,5){11.20}}
\put(24.00,14.67){\line(3,5){10.60}}
\put(20.33,26.33){\line(1,0){5.67}}
\put(31.00,26.33){\line(1,0){4.00}}
\put(22.67,21.33){\line(1,0){5.33}}
\end{picture}
\end{center}
\vspace*{.3mm}  

Probably the same
situation can occur with $N_f\geq3$. Finally, note that
near the origin, when D 2-brane is close to D 4-brane, in M theory picture 
some membranes become zero and then we will have tensionless strings in 
the worldvolume of 5-brane. This is similar to the case of $N=2$ $D=4$
gauge theory. So one can expect that there is a dual description similar
to the $N=2$ as speculated in\cite{DS}.

After completion of this work, I received the paper\cite{HB} which 
covers the same material as in this work
\vspace*{5mm}

I would like to thank F. Ardalan, M. M. Sheikh Jabbari and C. Vafa 
for useful comments.

\end{document}